\documentclass[sigconf]{acmart}
\usepackage{booktabs} 

\usepackage{tcolorbox}
\usepackage{subcaption}
\usepackage{tabularx}
\usepackage{array}
\usepackage{colortbl}
\usepackage{comment}
\tcbuselibrary{skins}
\usepackage{hyperref}  
\usepackage{hyperxmp}  

\usepackage{placeins}

\newcolumntype{Y}{>{\raggedleft\arraybackslash}X}

\tcbset{tab1/.style={fonttitle=\bfseries\large,fontupper=\normalsize\sffamily,
colback=white!10!white,colframe=black!75!black,colbacktitle=Salmon!40!white,
coltitle=black,center title,freelance,frame code={
\foreach \n in {north east,north west,south east,south west}
{\path [fill=black!75!black] (interior.\n) circle (0.5mm); };},}}

\tcbset{tab2/.style={enhanced,fonttitle=\bfseries,fontupper=\normalsize\sffamily,
colback=white!10!white,colframe=black!50!black,colbacktitle=Salmon!40!white,
coltitle=black,center title}}

\usepackage[subrefformat=parens]{subcaption}
\begin{document}

\title{Super-Resolution GPS: Restoring High-Precision Mobility Data}
\title{Restoring Super-High Resolution GPS Mobility Data}
\author{Haruki Yonekura}
\affiliation{
    \country{Japan}
    \institution{Osaka University}
}
\email{h-yonekura@ist.osaka-u.ac.jp}

\author{Ren Ozeki}
\affiliation{
    \country{Japan}
    \institution{Osaka University}
}
\email{r-ozeki@ist.osaka-u.ac.jp}

\author{Hamada Rizk}
\affiliation{%
    \country{Osaka University,  Japan}
    \institution{Tanta University, Egypt}
}
\email{hamada\_rizk@ist.osaka-u.ac.jp}

\author{Hirozumi Yamaguchi}
\affiliation{
    \country{Japan}
    \institution{Osaka University}
}
\email{h-yamagu@ist.osaka-u.ac.jp}

\begin{abstract}
This paper presents a novel system for reconstructing high-resolution GPS trajectory data from truncated or synthetic low-resolution inputs, addressing the critical challenge of balancing data utility with privacy preservation in mobility applications. The system integrates transformer-based encoder-decoder models with graph convolutional networks (GCNs) to effectively capture both the temporal dependencies of trajectory data and the spatial relationships in road networks. By combining these techniques, the system is able to recover fine-grained trajectory details that are lost through data truncation or rounding, a common practice to protect user privacy.
We evaluate the system on the Beijing trajectory dataset, demonstrating its superior performance over traditional map-matching algorithms and LSTM-based synthetic data generation methods. The proposed model achieves an average Fréchet distance of 0.198 km, significantly outperforming map-matching algorithms (0.632 km) and synthetic trajectory models (0.498 km).
The results show that the system is not only capable of accurately reconstructing real-world trajectories but also generalizes effectively to synthetic data. 
These findings suggest that the system can be deployed in urban mobility applications, providing both high accuracy and robust privacy protection.
\end{abstract}





\maketitle

\section{Introduction}

Human mobility trajectory data is rapidly becoming a cornerstone in various sectors, powering applications that are integral to the efficient operation of smart cities. From optimizing transportation systems and traffic management to supporting urban planning and ensuring effective crisis management during emergencies, mobility data underpins the design and deployment of data-driven services. As cities grow increasingly complex, the accurate tracking and analysis of movement patterns of large populations can significantly improve urban decision-making, resource allocation, and safety. However, the widespread availability and utilization of such data also present substantial challenges related to data quality, privacy, storage, and network efficiency.

At the core of these challenges is the vast amount of trajectory data generated by GPS-enabled devices, such as smartphones, vehicles, and IoT sensors. This data is frequently used to model and predict human movement patterns for applications such as ride-sharing services, public transit optimization, and crowd control in emergencies. Yet, despite its usefulness, raw GPS data often requires rounding or compression to meet practical constraints. These include preserving user privacy, reducing storage and transmission costs, and mitigating issues related to GPS signal noise and sensitivity, especially in urban areas with dense infrastructure. Such transformations, while necessary, result in a reduction in the spatial and temporal resolution of the data, limiting its effectiveness for fine-grained analysis.

The trade-off between compression and the retention of critical trajectory information is a long-standing problem. Traditional lossy compression methods have been developed to reduce the size of GPS data, but they tend to obscure key features such as rapid changes in direction, sharp turns, or short-distance deviations. These dynamics are essential for understanding specific mobility behaviors, such as vehicle movements in congested areas or pedestrian paths in crowded environments. Furthermore, the compression process often removes intricate details that could be vital for applications like precise traffic flow analysis, emergency evacuation planning, or identifying anomalous behavior patterns in real-time \cite{10.1007/s10707-021-00434-1, 10.1007/s10707-013-0184-0}.

Despite these advancements in data compression, research on methods to recover or enhance low-resolution GPS data that have been compressed or rounded for storage or privacy reasons remains limited. 
In our preliminary work \cite{gotoprivacy, 10.1145/3589132.3625651, ozeki2024privacy}, we explored methods to accurately predict taxi demand using distributed GPS data, representing an approach toward privacy-preserving data analysis by reducing the need for centralized data collection.
The ability to reconstruct high-resolution GPS data from low-resolution counterparts holds immense potential for enhancing the usability of mobility services in smart cities. For instance, reconstructing fine-grained movement patterns from aggregated data could significantly improve applications that rely on high precision, such as dynamic navigation systems, autonomous vehicle guidance, or smart grid energy optimization. More importantly, this capability could serve as a privacy-preserving mechanism, allowing detailed analyses to be performed without exposing individuals' exact locations.

This reconstruction problem shares similarities with challenges in computer vision, where high-resolution images are often reconstructed from low-resolution inputs through sophisticated models such as super-resolution techniques \cite{yu2024review}. In the context of mobility trajectory data, the goal is not just to recover spatial resolution but also to preserve the complex temporal dependencies and contextual information, such as the underlying road network or movement patterns of the surrounding environment. Current solutions fall short in this regard, as they primarily focus on minimizing data size rather than intelligently restoring it for specific downstream tasks.

Motivated by these limitations, in this paper, we propose a novel system that reconstructs truncated or low-resolution GPS trajectory data to recover its utility and precision. Our approach integrates two powerful components: a transformer-based encoder-decoder model to handle the spatio-temporal characteristics of the GPS trajectory data and a graph neural network (GNN) to incorporate flexible road network information.
Transformers, known for their ability to capture long-range dependencies in time-series data, are particularly well-suited for handling the spatio-temporal nature of GPS trajectories. In our system, the transformer-based model serves as the foundation for learning and encoding the temporal dynamics and spatial correlations inherent in the trajectory data. By modeling the relationship between successive GPS points over time, this approach captures both short-term fluctuations and long-term trends in mobility patterns, enabling accurate trajectory reconstruction.
In addition to the transformer, we employ a graph neural network (GNN) that processes the underlying road network information. Urban road networks are inherently structured as graphs, where nodes represent intersections and edges represent roads. By leveraging GNNs, we can embed this graph structure into the trajectory reconstruction process, ensuring that the reconstructed GPS points align with realistic road constraints and follow plausible mobility routes. This fusion of road network information with GPS data allows our system to generate high-resolution trajectories that are not only spatially accurate but also contextually aware of the road environment.

By integrating these components, our system effectively reconstructs high-resolution trajectories from compressed or rounded GPS data, mitigating the loss of crucial information during the compression process. This approach addresses the key limitations of existing methods, enabling more robust and reliable trajectory-based services in the context of smart city applications. Ultimately, this research contributes to the growing demand for data-efficient and privacy-preserving solutions, while ensuring the recovery of high-precision mobility data that is essential for next-generation urban systems.

\color{black}

\section{Related Work}
\subsection{Privacy-preserving method on human mobility data}

Spatio-temporal data plays a crucial role in training data-driven models for various applications, including taxi demand prediction. However, gathering such data is expensive and can expose sensitive information, thereby posing risks to user privacy. Consequently, numerous studies have been devoted to developing techniques that safeguard the privacy of data and machine learning models.

Several studies \cite{4417165, 10.1145/1869790.1869846} have concentrated on protecting user privacy by introducing dummy data into mobility datasets. 
The approach proposed in \cite{4417165} introduces false location points into the trajectory in either a random or rotational fashion.
To create more realistic trajectories, the authors of \cite{10.1145/1869790.1869846} proposed generating dummy locations by constraining the user’s movement. 
This technique enhances anonymity within a specified spatial area while maintaining the plausibility of the trajectory. 
However, it has limitations, such as when an attacker possesses prior knowledge of the user's lifestyle, they could potentially infer the user’s interests or actual location \cite{HEMKUMAR20201291}.

In addition to these approaches, synthesis-based methods for privacy protection have been introduced, which involve replacing the original mobility data with synthetic data \cite{LSTM-trajGAN, 4497446, zhu2023diffusion}. 
One of the key concepts in this context is k-anonymity, which guarantees privacy by ensuring that there are at least $k$ users with similar characteristics, thus preventing an attacker from narrowing down the potential users to fewer than $k$, even when attempting to identify a user based on specific traits. 
To achieve k-anonymity in mobility data, the method proposed in \cite{4497446} incorporates uncertainty into the location data. 
To address privacy concerns more effectively, differential privacy has emerged as a leading technique for anonymization and privacy preservation \cite{10.1145/2484838.2484846,10.1145/3423165}. This is a mathematical framework that adds noise to the data, ensuring a certain degree of privacy protection. Several techniques, such as CNoise and SDD \cite{10.1145/2484838.2484846}, have been introduced to apply differential privacy to mobility trajectory data by adding noise to guarantee privacy. The added noise and the corresponding level of privacy have been mathematically validated, making this approach more robust than others.
For creating realistic synthetic mobility data, DiffTraj \cite{zhu2023diffusion} generates a synthetic location dataset using a diffusion model. 
LSTM-trajGAN and recent pioneering work\cite{ozeki2022sharing, ourMDMpaper, yonekura2023stm}, on the other hand, generates synthetic mobility data using LSTM and replaces the original dataset with this synthetic version. 
Since LSTM-trajGAN employs a GAN-based approach to train the LSTM generator, the resulting synthetic trajectories are expected to be highly realistic. 
Nevertheless, these methods do not fully guarantee the privacy and utility of the generated data.
\color{black}

\subsection{GPS trajectory reconstruction}
Research in addressing low-sampled or incomplete trajectories has been ongoing for over a decade.
The significant demand for high-resolution trajectories has driven efforts to insert additional points between consecutive trajectory points. 
Various terms, including trajectory interpolation, completion, recovery, and reconstruction commonly refer to this process.

One common approach to trajectory reconstruction is map matching, as demonstrated in studies like \cite{newson2009hidden, yuan2010interactive, lou2009map}.
These methods often employ models like the hidden Markov model or rely on a voting process among sampled points.
Other approaches \cite{8834829, li2016knowledge} leverage historical trajectory data and road network information to recover missing trajectories.
The authors of \cite{li2016knowledge} focus on junctions in urban regions, extracting a basic trajectory skeleton from sparse data and using a clustering framework to reconstruct incomplete GPS trajectories.
The authors of \cite{8834829} use LSTM-based autoencoder architecture for recovery from sub-trajectory to the whole trajectory and use Kalman Filter to decrease the noise and calibrate trajectory estimation.
KAMEL\cite{musleh2023kamel} adapts the BERT concept to realize trajectory imputation without using map geometry.
They partition the area and regard a cell as a word so that the data can be processed through a BERT-like architecture.
By introducing the spatial constraints, they make it possible to impute trajectories.
Additionally, there are studies that do not rely on map information for trajectory reconstruction or imputation, such as \cite{10.1145/2939672.2939833, Ruan_Long_Bao_Li_Yu_Li_Liang_He_Zheng_2020, 10.1145/3582427}. 
Among them, INGRAIN\cite{10.1145/3582427} employs the attention mechanism of transformer architectures to impute missing data.

In comparison to these existing approaches, our proposed system integrates transformers for understanding GPS time-series data and graph neural networks for spatial data interpretation, offering a more comprehensive solution to trajectory reconstruction.

\color{black}

\begin{figure*}[!tbp]
    \centering
    \includegraphics[width=0.9\linewidth]{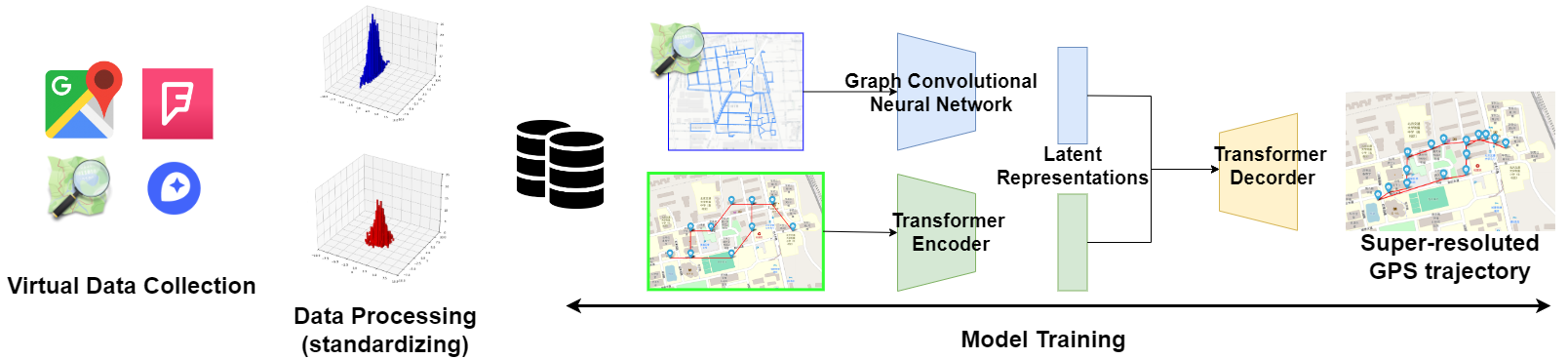}
    \caption{System overview.}
    \label{fig:system_overview}
\end{figure*}


\section{System details}

The system is designed to reconstruct high-resolution GPS trajectory data from low-resolution, privacy-preserved inputs. As shown in Figure \ref{fig:system_overview}, it consists of three main modules: virtual data collection, data processing, and model training. These components work in tandem, utilizing advanced deep learning methods such as Graph Convolutional Networks (GCNs) and Transformer models to accurately reconstruct trajectory data while taking into account both the spatial and temporal dimensions.
\color{black}

\subsection{Virtual Data Collection}
The first step in our system involves the generation of virtual data that captures general mobility patterns within a specified geographical region. This is done by randomly selecting points within the target region and using publicly available tools, such as routing algorithms, to calculate the shortest paths between these points. The output of this process is a set of trajectories, and each is represented as a sequence of latitude and longitude pairs. These trajectories are critical for simulating real-world mobility patterns that can be used to train the super-resolution model.

In parallel, we extract the graph structure of the road network associated with the generated trajectories. The road network is modeled as a graph, where nodes represent intersections and edges represent the haversine distance between adjacent intersections. Each node in the graph is annotated with the latitude and longitude of the corresponding intersection, and the edges are weighted based on the geospatial distance between intersections. This graph structure provides crucial contextual information about the layout of the area, enabling the model to incorporate local road network constraints during trajectory reconstruction. By allowing users to generate data for any region with adjustable parameters, the system remains flexible and adaptable to different geographical contexts, enabling scalability across various application domains.
\color{black}

\subsection{Data Processing}\label{sec:data_processing}
Once the virtual data is generated, it undergoes several preprocessing steps to ensure compatibility with the model and simulate real-world scenarios where GPS data is often altered to preserve privacy or reduce storage overhead. This step involves artificially modifying the virtual trajectories by applying techniques such as rounding the latitude and longitude values, adding noise, or implementing spatial cloaking methods. These modifications mimic common practices used in privacy-preserving GPS data applications, where high-resolution data is intentionally degraded to protect user identity or to optimize storage and transmission efficiency.

The modified (or low-resolution) trajectories are then used as inputs to the model, while the corresponding original (or high-resolution) trajectories serve as the ground truth for training. The goal of the model is to learn how to accurately reconstruct the original high-resolution trajectory from the degraded version. To facilitate learning, we normalize all latitude and longitude values using z-score normalization. This ensures that the data fed into the model is standardized across regions, preventing the model from being biased toward specific geographic scales. Additionally, the distances between intersections (represented as edges in the road network graph) are transformed by taking their inverse, which enhances the propagation of information in the GCN by prioritizing shorter, more relevant distances between nodes. This step helps the model focus on nearby intersections, where small variations in trajectory data can have significant implications for the reconstruction process.
\color{black}

\subsection{Model Training}
The core of the system lies in its two deep learning components: the Graph Convolutional Network and the Transformer-based encoder-decoder architecture. These components work together to process both the road network information and the spatio-temporal trajectory data, ultimately generating a high-resolution output.

\textbf{GCN for road network embedding:}~
Road networks are highly variable in their structure, with differences in the number of intersections, the density of roads, and the layout of cities or regions. Traditional machine learning models struggle to handle this complexity due to their reliance on fixed-length inputs. To address this, we treat the road network as a graph, where nodes correspond to road intersections, and edges represent the connections between them. The GCN allows us to learn a rich embedding of this graph, capturing the structural relationships between nodes while considering the geospatial proximity of the roads.

For each node (intersection), the latitude and longitude serve as features, while the inverse haversine distance between connected nodes serves as edge weights. By using platforms such as OpenStreetMap, we can obtain detailed road network data for any region, allowing the model to adapt to different geographic areas. During the training process, the GCN computes a node embedding for each intersection, which encodes the local road network information relevant to the trajectory reconstruction task. To optimize memory efficiency and computation time, we limit the graph to the area surrounding the trajectory, instead of using the entire region covered by the dataset. This localized approach improves the message-passing mechanism within the GCN, ensuring that the model focuses on the most relevant parts of the road network.


\textbf{Transformer Encoder for Trajectory Embedding:}~

The transformer encoder is responsible for modeling the sequential and temporal aspects of the GPS trajectory data. GPS trajectories are inherently time-series data, where each point is defined not only by its spatial coordinates but also by its temporal order. To capture these dependencies, we use a multivariate transformer model that processes the standardized latitude, longitude, and timestamp values for each point in the trajectory.

Since trajectories vary in length, the model dynamically adjusts by padding shorter sequences and masking irrelevant inputs to ensure consistency during training. This allows the transformer to focus on meaningful sections of the trajectory, avoiding bias from extraneous padding. The transformer encoder learns a representation of the trajectory that captures both local variations (such as short-term changes in direction) and global patterns (such as long-term trends in movement), enabling the model to reconstruct fine-grained details in the GPS data.

\color{black}
\textbf{Transformer Decoder for Super-Resolution:}~


The transformer decoder serves as the final component of the system, where the goal is to enhance the resolution of the input trajectory. The decoder combines the embeddings generated by the GCN (which captures road network information) and the transformer encoder (which captures spatio-temporal dynamics) to produce a high-resolution trajectory. This is achieved by mapping the low-resolution input to a finer-grained output that closely resembles the original trajectory.

To transform the standardized outputs back into actual latitude and longitude values, we apply reverse normalization using the mean and variance of the original data. This ensures that the reconstructed trajectory corresponds to real-world GPS coordinates, preserving its spatial fidelity. We use Soft Dynamic Time Warping (SoftDTW) as the loss function, which is particularly effective for measuring the similarity between time-series data. SoftDTW accounts for potential temporal misalignments between the reconstructed trajectory and the ground truth, ensuring that the model accurately captures the underlying movement patterns even when there are small deviations in timing. The loss function is minimized through backpropagation, enabling the model to iteratively refine its predictions and improve the accuracy of the reconstructed trajectory.

Through the combined use of GCNs and Transformers, our system is able to effectively reconstruct high-resolution GPS trajectory data from degraded inputs, addressing the challenges posed by privacy-preserving modifications and data compression. The integration of road network information ensures that the reconstructed trajectories align with real-world geography, while the transformer model captures the nuanced spatio-temporal dependencies that are essential for accurate mobility modeling.
\color{black}

\section{Evaluation}
The evaluation of the proposed system was conducted to assess its ability to reconstruct high-resolution GPS trajectories from low-resolution. This section provides a detailed explanation of the experimental settings, evaluation metrics, and the results obtained. The experiments focus on the performance comparison between the proposed model and traditional methods for trajectory reconstruction, specifically map-matching algorithms.

\subsection{Experimental Settings}

\begin{figure}
    \centering
    \includegraphics[width=0.95\linewidth]{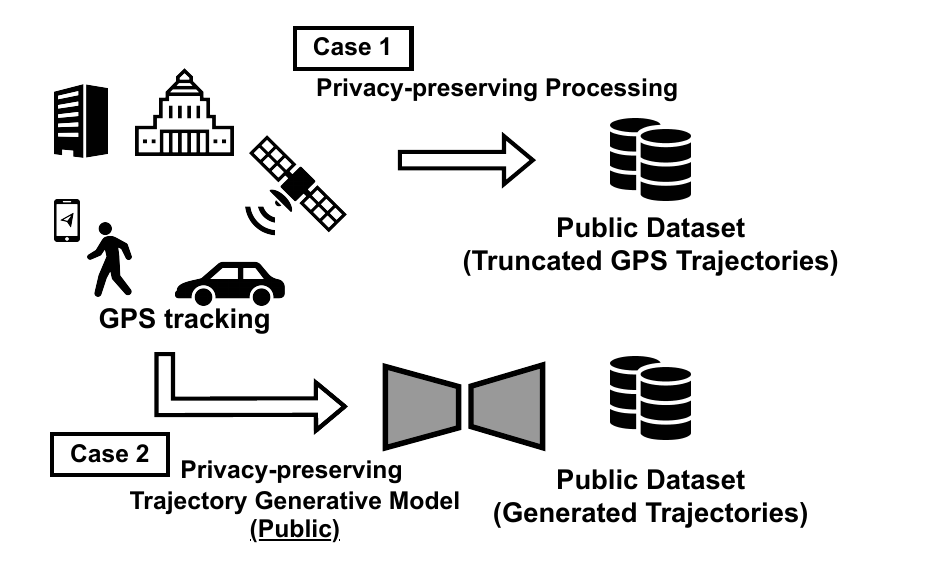}
    \caption{Data publication scenarios.}
    \label{fig:attack_scenario}
\end{figure}



\begin{figure*}[t]
    \centering
    \begin{subfigure}{0.35\textwidth}
        \centering
        \includegraphics[width=\linewidth]{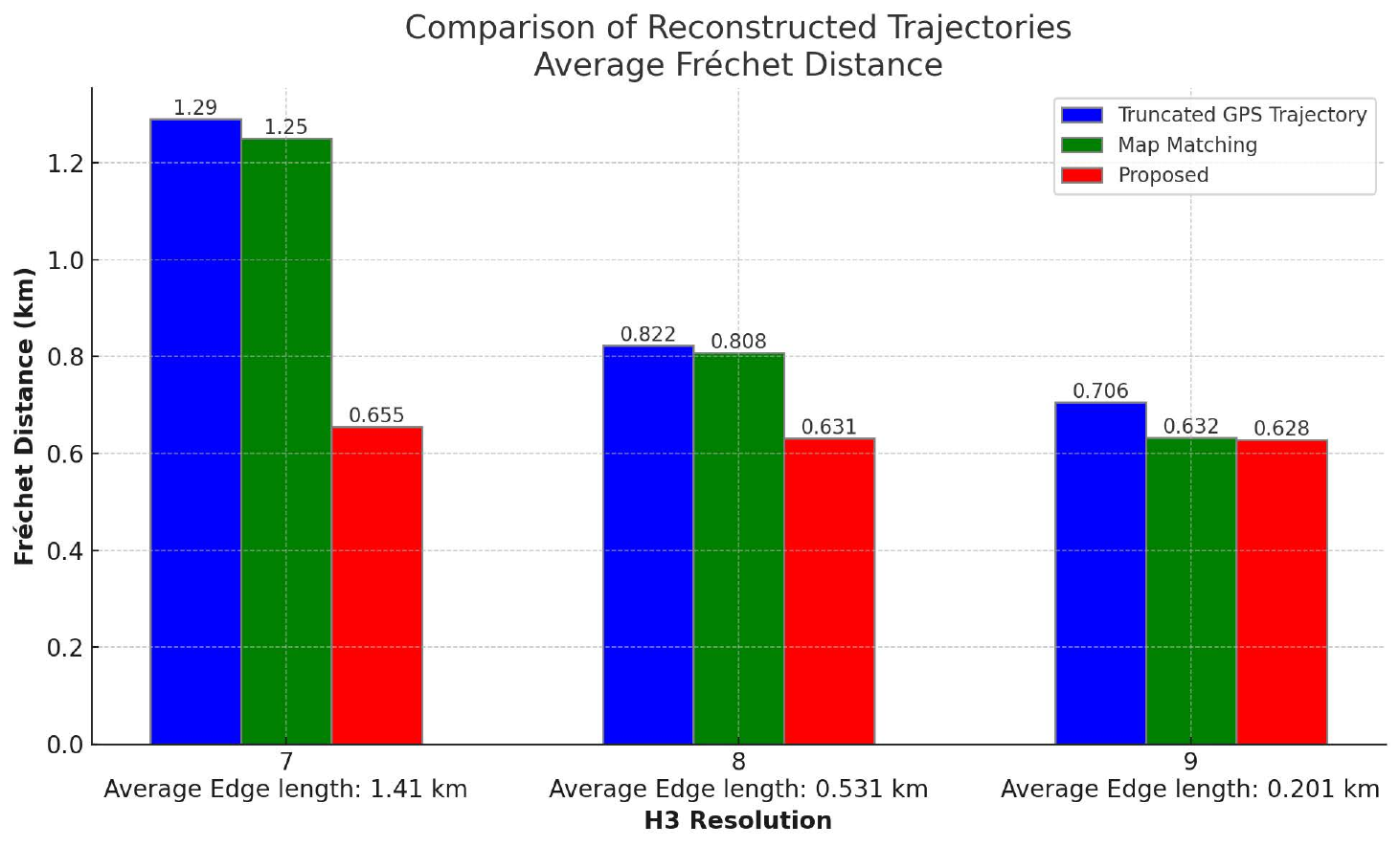}
        \caption{Distance from actual trajectory[km].}
        \label{fig:reconstruct_statistic_result}
    \end{subfigure}%
    \hfill
    \begin{subfigure}{0.35\textwidth}
        \centering
        \includegraphics[width=\linewidth, height=4cm]{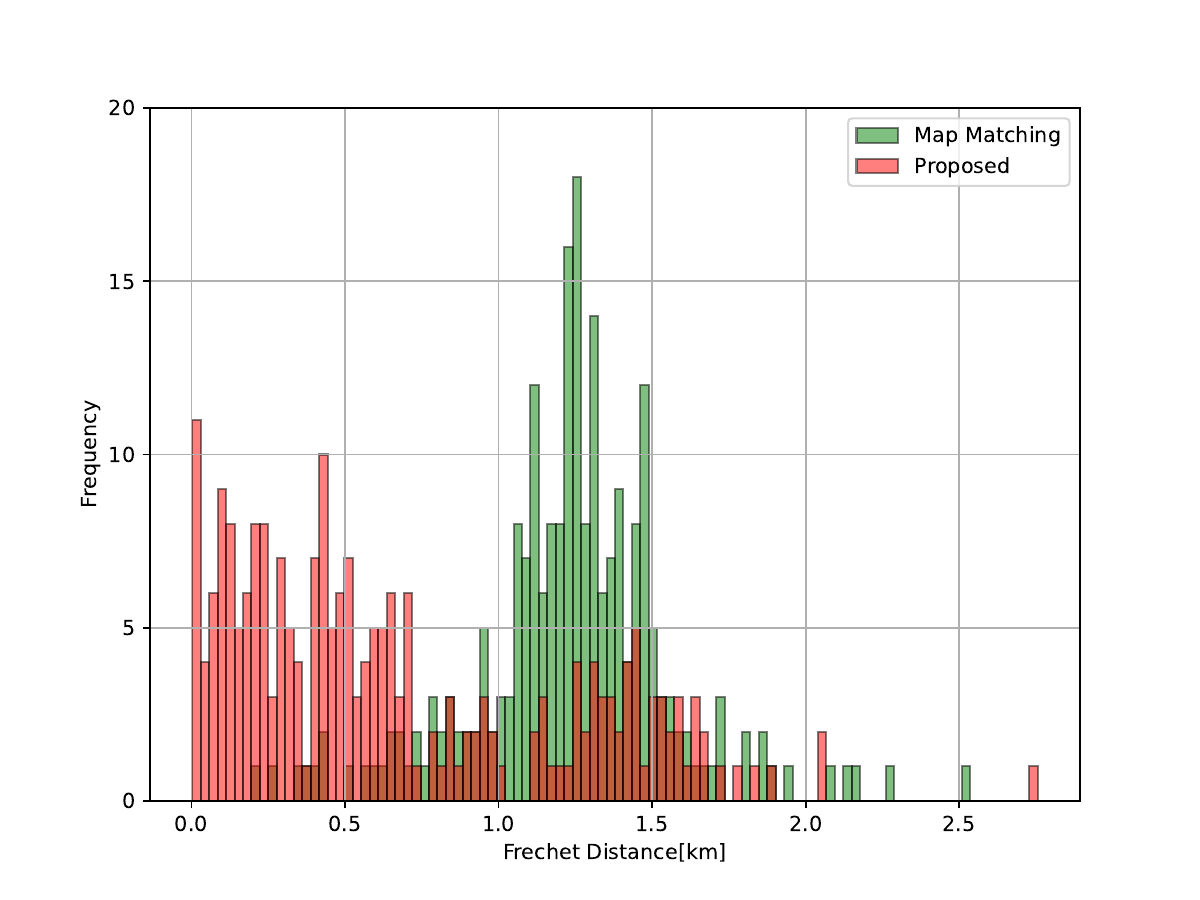}
        \caption{Histogram of the error distance (H3 resolution is set as 7).}
        \label{fig:reconstruct_statistic_histgram}
    \end{subfigure}%
    \hfill
    \begin{subfigure}{0.27\textwidth}
        \centering
        \includegraphics[width=\linewidth, height=4cm]{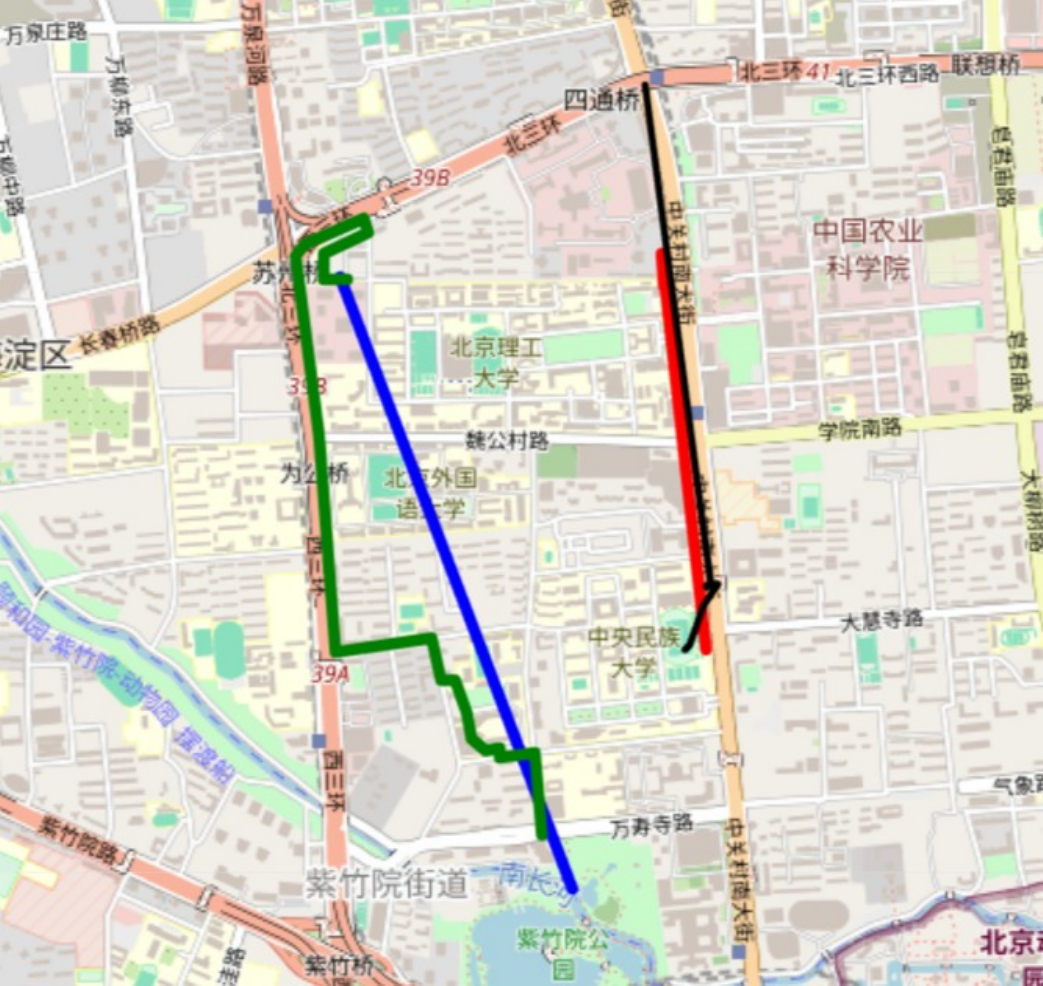}
        \caption{One of the samples of super-resolution trajectory.}
        \label{fig:generated_result}
    \end{subfigure}
    \caption{Comparison of reconstructed trajectories, error histograms, and visualizations.}
\end{figure*}

To evaluate the proposed system, we utilized the Beijing trajectory dataset \cite{zheng2011geolife}, a large-scale public dataset containing real-world human mobility data. From this dataset, we selected trajectories with sequence lengths not exceeding 128 points. We generated a total of 25,769 virtual GPS trajectories within the Beijing region to train our model. The synthetic data generation ensures that the model can learn region-specific mobility patterns while being tested on real-world trajectory data. This approach allows us to control the characteristics of the data used for training, such as sequence length and density, making it adaptable to different urban scenarios.

For privacy preservation, we applied rounding to the GPS values in the dataset using the H3 indexing system \cite{wozniak2021hex2vec}. Each GPS point was snapped to the center of its corresponding H3 hexagonal cell, ensuring that the spatial resolution of the data was reduced. This truncation simulates real-world practices where GPS data is often obfuscated for privacy protection. \footnote{H3 is a geospatial indexing system that divides the globe into hexagonal cells. Each cell is assigned a unique identifier, and by snapping GPS points to the nearest cell center, we reduce the granularity of the data, preserving privacy while introducing spatial errors.}

The evaluation was conducted in two scenarios:
\begin{itemize}
    \item  \textbf{Scenario 1:} The GPS trajectory data is directly rounded and published without any additional processing.
\item \textbf{Scenario 2:} A synthetic trajectory dataset is generated using a model trained on real data, such as the LSTM-based TrajGAN model \cite{ozeki2023balancing}. The synthetic data is made publicly available, and the data users can employ the trained model to generate more synthetic trajectories for further analysis.
\end{itemize}

These two scenarios allow us to assess how the proposed system performs in both simple truncation cases and more complex generative cases. Scenario 2 introduces a more challenging task where synthetic data needs to be transformed into high-resolution outputs.

For comparison, we employed a traditional map-matching algorithm \cite{newson2009hidden}, which aligns low-resolution GPS points to the nearest roads in a map database, aiming to reconstruct the most likely trajectory.

\color{black}

\subsection{Result}

The evaluation of the proposed system’s performance was based on the Fréchet distance, a widely recognized metric for measuring the similarity between two curves by accounting for both spatial and temporal alignment. This metric is particularly well-suited for trajectory data because it captures not only the spatial deviations between paths but also their timing, making it an ideal tool for assessing the fidelity of reconstructed trajectories relative to the original ones. Lower Fréchet distances indicate a closer match between the reconstructed and original high-resolution trajectories.

The visual results in Figure \ref{fig:reconstruct_statistic_result} provide a clear comparison of the original, truncated, and reconstructed trajectories. The proposed system demonstrates exceptional performance in restoring detailed movement paths.
In Scenario 1, where GPS points were rounded to reduce spatial resolution, our model was able to reconstruct trajectories that follow the original path with high accuracy, effectively recovering fine-grained details lost during truncation.

A more detailed comparison is shown in Figure~\ref{fig:reconstruct_statistic_histgram}, where the truncated trajectory, map-matched trajectory, and reconstructed trajectory are visually examined. The proposed system clearly recovers finer details, even from heavily truncated or synthetic data, which map-matching fails to restore accurately. This demonstrates the system's ability to infer complex movement patterns that traditional methods overlook, particularly in scenarios where GPS points have been heavily degraded or artificially generated.
Further analysis of the Fr\'{e}chet distance is provided in the histogram in Figure \ref{fig:reconstruct_statistic_histgram}, which shows that 85.7\% of the reconstructed trajectories fall within a very narrow error margin. This high success rate underscores the system’s ability to handle a wide range of truncation levels effectively, further validated across various H3 resolution levels. The model's flexibility in maintaining performance at different spatial granularities highlights its adaptability, making it suitable for diverse real-world applications, from low-resolution data sources to heavily obfuscated trajectories.

In Figure \ref{fig:generated_result}, the visual comparison between the truncated trajectory, map-matched trajectory, and the reconstructed trajectory clearly shows that the proposed model more accurately recovers the fine-grained details of the movement, even from heavily truncated or synthetic data.



The performance metrics, summarized in Table~\ref{tab:summary_lstm_sutoencoder_generated}, provide further evidence of the model’s strength. Our proposed system achieves an average reconstruction error of 0.198 km, which is a significant improvement over the map-matching algorithm’s average error of 0.632 km.
This result demonstrates superior generalization capabilities. The substantial reduction in reconstruction error suggests that the model can handle both real and synthetic data effectively, improving upon the initial inaccuracies introduced during the generative process.

The proposed system’s significant improvement in Fr\'{e}chet distance across both scenarios confirms the advantages of integrating Transformer and GCN models for GPS trajectory reconstruction. The Transformer captures long-range dependencies and temporal patterns, while the GCN effectively models spatial relationships in the road network. This combination allows the system to restore fine-grained mobility patterns with precision, making it particularly valuable for applications in urban environments, such as transportation planning, traffic flow optimization, and emergency response.

The results clearly demonstrate that, by leveraging the spatial and temporal dimensions of trajectory data, the proposed system offers a robust solution to the challenges of GPS trajectory reconstruction, particularly in cases where data truncation or obfuscation is necessary for privacy preservation. The ability to recover accurate, high-resolution trajectories from heavily truncated data positions the system as a powerful tool for a wide range of urban and mobility applications, ensuring both privacy protection and data utility.

\begin{table}
    \centering
    \caption{LSTM autoencoder trajectories summary.}
    \begin{tabular}{c|c} \hline
       Trajectory & Distance\\ \hline
       LSTM & 0.498 km \\
       Map matching trajectory & 0.632 km \\
       Proposed system & 0.198 km \\\hline
    \end{tabular}
    \label{tab:summary_lstm_sutoencoder_generated}
\end{table}

\color{black}

\section{Conclusion}
In this work, we proposed and developed a robust system for reconstructing high-resolution GPS trajectory data from low-resolution, privacy-preserved inputs. By integrating transformer-based models and GCNs, the system effectively captures the temporal dynamics of human movement while incorporating the spatial structure of road networks. This hybrid approach enables the model to restore fine-grained trajectory data that would otherwise be lost due to truncation, rounding, or the use of synthetic data generation techniques.
The evaluation of the system using the Beijing trajectory dataset confirmed its significant performance advantages over traditional map-matching algorithms and LSTM-based synthetic trajectory models. The proposed model consistently produced lower Fréchet distances, demonstrating its ability to handle both real and synthetic data with high accuracy. The results indicate that the system is particularly well-suited for urban applications that require precise mobility data, such as transportation planning, emergency response, and smart city services, while maintaining strong privacy protections.
Future work may explore the system's adaptability to other geographical regions and the integration of additional privacy-preserving techniques to further enhance the balance between data utility and user privacy.


\bibliographystyle{ACM-Reference-Format}
\bibliography{_ref.MonoFi}

\end{document}